\documentclass[aps,preprint,amssymb]{revtex4}

\usepackage{epsfig}

\begin{document}
\renewcommand{\thefootnote}{\fnsymbol{footnote}}
\renewcommand{\theequation}{\arabic{section}.\arabic{equation}}

\renewcommand{\vec}[1]{\mathbf{#1}}
\newcommand{\eqref}[1]{(\ref{#1})}

\title{Coupling of ion and network dynamics in lithium silicate glasses: a computer study}

\author{Magnus Kunow}
 \email{kunow@uni-muenster.de}
\author{Andreas Heuer}
 \email{andheuer@uni-muenster.de}

\affiliation{Institut f\"{u}r Physikalische Chemie and Sonderforschungsbereich 458,
Westf\"{a}lische Wilhelms-Universit\"{a}t, Corrensstra{\ss}e 30, D-48149 M\"{u}nster, Germany
}

\date{\today}

\begin{abstract}
\noindent We present a detailed analysis of the ion hopping
dynamics and the related nearby oxygen dynamics in a lithium meta
silicate glass via molecular dynamics simulation. For this purpose
we have developed numerical techniques to identify ion hops and to
sample and average dynamic information of the particles involved.
This leads to an instructive insight into the microscopic
interplay of ions and network. It turns out that the cooperative
dynamics of lithium and oxygen can be characterized as a sliding
door mechanism. It is rationalized why the local network
fluctuations are of utmost importance for the lithium dynamics.
\end{abstract}

\maketitle

\section{Introduction}
\label{intro} As a well known characteristic feature of ion
conducting glasses the dynamics of ions and network atoms are
strongly decoupled at low temperatures and in particular below the
glass transition temperature $T_\mathrm{g}$
\cite{howell74,angell86,heuer02}. Although by definition for
temperatures $T<T_\mathrm{g}$ the network dynamics is mainly
limited to localized fluctuations, molecular dynamics simulations
of alkali silicate systems with an immobilized network give clear
evidence that the network dynamics has a significant impact on the
ion dynamics \cite{angell81, sunyer03, habasaki04, heuer05}. A
similar behavior has already been predicted by Anderson and Stuart
who have assumed that the opening of diffusion pathways between
adjacent ionic sites through the silicate matrix supplies a major
contribution to the activation energy of the ion diffusion in
alkali silicate glasses \cite{anderson54,martin86}. In recent work
it has been proposed that at least for mixed-alkali systems a
strong coupling of alkali and network dynamics should be present
to rationalize experimental findings \cite{cmmc}. Furthermore for
models of ion dynamics it is particularly important to know the
nature of the network relaxation if an ion is leaving its site
\cite{ingramnew}.

In \cite{heuer05} we have recently shown that for a lithium
silicate system with an artificially immobilized silicate network
the lithium dynamics changes in many aspects compared to a system
with unchanged network mobility. A dramatic slowing down of the
ion dynamics is observed, leading to an increase of the activation
energy for the lithium diffusion by a factor of 1.7. Furthermore,
it has turned out to be the immobilization of the oxygen particles
that contributes most to this behavior. Analysis of suitable
three-time-correlation functions has shown that forward-backward
correlations of the lithium dynamics is much more pronounced.
Thus, the minima of the potential landscape inside the fixed
network structure act on the ions like an effective harmonic
potential for significantly larger displacements of the ions than
this would be the case inside a network carrying out its natural
dynamics. Furthermore, by immobilizing the network species the
lithium dynamics gets more heterogeneous. Recently, also the
effect of immobilization of a few ions on the the dynamics of the
remaining ions has been studied \cite{habasaki_new}.

Another important ingredient for the ion dynamics through the
glass matrix concerns the availability of a vacant site in the
neighbourhood of an ion. This problem gains particular importance
if the number density of ions is close the number density of
sites. From molecular dynamics simulations by Lammert \textit{et
al.}, Habasaki \textit{et al.} and Vogel it has turned out that in
the lithium silicate system investigated in this work the number
density of sites is only few percent larger than the one of ions
\cite{lammert03, habasaki04,vogel:2004}. This result is consistent
with the previous observation of ion channels \cite{horbach:2002}.
Thus, for most ions there is no adjacent vacant site to which it
might hop. These ions have to wait for an adjacent ion to perform
a hop leaving a vacant side before they can perform a hop on their
own. Thus, a macroscopic charge transport requires some
cooperative hopping mechanism. Such correlated subsequent hops
have been observed by Cormack \textit{et al.} simulating a sodium
silicate system \cite{cormack02}. For lithium silicate correlated
subsequent hops have been found by Habasaki \textit{et al.},
especially for ions with high mobility \cite{habasaki95-2,
habasaki96, habasaki97, habasaki99}. They traced this behavior
back to the prevention of a correlated backward hop due to the
re-occupation of the initial site.

The goal of this work is to identify ion hops and sample the
relevant dynamic information. Our analysis goes beyond the
observation of individual hopping scenarios. By calculating
averages over many different hopping events (of the order $10^5$)
we are able to make statistically relevant statements about the
nature of the dynamics.

We have organized this paper as follows. In sec.\ \ref{tech} we
describe the technical aspects of the simulation and present the
details of our analysis. The results are shown in sec.\
\ref{Results}. Finally, conclusions are drawn and discussed in
sec.\ \ref{concl}.

\section{Technical aspects}
\label{tech}
\subsection{Simulation}
\label{sim}
The potential energy of the lithium silicate system is chosen to be the sum of a Buckingham and a Coulomb pair potential:
\begin{equation}\label{pot1}
U_{ij}(r_{ij}) =\frac{1}{4\pi\epsilon_0} \frac{q_i q_j}{r_{ij}} - \frac{C_{ij}}{r^6_{ij}} + A_{ij} \exp (-B_{ij} r_{ij}).
\end{equation}
The indices $i$ and $j$ denote the species lithium, oxygen, and
silicon, respectively. The potential parameters are listed in
previous work by Banhatti and Heuer \cite{banhatti01}. They are
based on \textit{ab initio} calculations by Habasaki \textit{et
al.} \cite{habasaki90,habasaki92,habasaki93}. Applying a slightly
modified version 2.16e of the MOLDY software package
\cite{refson}, we have generated molecular dynamics (MD)
trajectories in the NVT ensemble, i.e. at constant volume. The
length of the elementary time step has been chosen to be 2 fs and
periodic boundary conditions have been used. We have simulated a
system of $N=1152$ particles at a density $\rho =
2.34\;\mathrm{g\,cm^{-3}}$ taken from experimental room
temperature data \cite{doweidar}. If the temperature is not
explicitly noted the presented data refers to $T=980\;\mathrm{K}$.

We have also modified the MOLDY software to compute only the
dynamics of a chosen subensemble of particles and keep the
positions of the other particles constant. Doing so, we are able
to take a configuration of our system at an arbitrarily chosen
time and continue the dynamics only for the ions so that the
network is artificially immobilized.

\subsection{Definition of lithium hops}
\label{define_hops} Qualitatively, a hop is identified as a
relatively large displacement during a relatively short time.
Thus, we take a minimum displacement $s_\mathrm{min}$ occurring
during a time interval $2\tau$ as the criterion for an ion hop. As
we are interested in hops between next neighbor positions the
value of $s_\mathrm{min}$ ought to be somewhat larger than the
first minimum distance $r_\mathrm{min}\approx1.5\;\mathrm{\AA}$ of
the self part of the van Hove correlation function and somewhat
less than the average hopping distance $d_0\approx
2.6\;\mathrm{\AA}$ \cite{heuer02}. The average time an ion resides
outside a lithium site has been found to be $500\;\mathrm{fs}$
\cite{lammert03}. The total time of the hop, also including the
time to leave the initial site and to enter its new site is
somewhat larger. Here we choose $\tau = 1$ ps. The actual choice
is uncritical (see below).

Let us assume $t_0$ to be the time when a hopping ion is crossing
the saddle point between two adjacent sites. To identify the
position $\vec{r}_\mathrm{sl}$ of this saddle as well as the
position $\vec{r}_\mathrm{init}$ of the initial lithium site and
$\vec{r}_\mathrm{end}$ of the lithium site after the hop, it has
turned out to be very useful to average the position of the ion at
times $t_0$, $t_0-\tau$, $t_0+\tau$, respectively, over a short
time. Since a priori $t_0$ is not known we define for these
averages:
\begin{eqnarray}
\label{r_sl}\vec{r}_\mathrm{sl}(t) &=& \frac{1}{2n_\mathrm{sl}+1}\sum_{i=-n_\mathrm{sl}}^{n_\mathrm{sl}}\vec{r}(t+it_\mathrm{slice}),\\
\label{r_init}\vec{r}_\mathrm{init}(t) &=& \frac{1}{2n_\mathrm{init}+1}\sum_{i=-n_\mathrm{init}}^{n_\mathrm{init}}\vec{r}(t-\tau +it_\mathrm{slice}),\\
\label{r_end}\vec{r}_\mathrm{end}(t) &=&
\frac{1}{2n_\mathrm{end}+1}\sum_{i=-n_\mathrm{end}}^{n_\mathrm{end}}\vec{r}(t+\tau
+it_\mathrm{slice}).
\end{eqnarray}
$t_\mathrm{slice}=20\;\mathrm{fs}$ is the time between subsequent
system configurations, stored during the MD run. Using
$n_\mathrm{sl}=5$ and $n_\mathrm{init}=n_\mathrm{end}=25$ the
position of the saddle point is averaged over $200\;\mathrm{fs}$,
and the positions of the lithium sites are averaged over
$1\;\mathrm{ps}$.

How to find an appropriate choice for $t_0$? For this purpose we
evaluate eqs.\ \eqref{r_sl} to \eqref{r_end} for all available
times $t$ and for every lithium ion. For an ion hop one expects:
\begin{equation}\label{condition1}
|\vec{r}_\mathrm{end}-\vec{r}_\mathrm{init}|\geq s_\mathrm{min}.
\end{equation}
For a single hopping event one typically finds a set of subsequent
points in time $t$ which do all fulfil the condition
\eqref{condition1}. We identify the exact time $t_0$ of the hop as
the time $t$ when the hopping lithium ion is half-way between the
initial site and the new site. Thus, the best value $t_0$
representing this particular ion hop is chosen by the criterion:
\begin{equation}\label{condition2}
\left|
|\vec{r}_\mathrm{end}(t)-\vec{r}_\mathrm{sl}(t)|-|\vec{r}_\mathrm{sl}(t)-\vec{r}_\mathrm{init}(t)|
\right| =\;\mathrm{min}.
\end{equation}

The average hopping distance $\left<
|\vec{r}_\mathrm{end}-\vec{r}_\mathrm{sl}| \right>$ is closest to
$d_0$ with the choice $s_\mathrm{min}=2\;\mathrm{\AA}$. Why is
$\tau = 1$ ps a good choice? An appropriate criterion for the
choice of $\tau$ is that the number of identified lithium hops is
a maximum. It turns out that the number of identified jumps is
basically constant for $0.7\;\mathrm{ps}\leq\tau\leq
1.3\;\mathrm{ps}$. Analyzing a simulation run of $20\;\mathrm{ns}$
length at $T=980\;\mathrm{K}$ with this method we find more than
130000 ionic hops.

\subsection{Coordinate Frame}
\label{frames} We wish to study the densities of the ions and the
network species (see section \ref{define_functions}) in the
temporal and spatial surrounding of each identified lithium hop.
To average over all hops we need to define frames of reference
that provide a certain sensitivity to the characteristic changes
we are interested in. First, we define a time scale relative to
$t_0$:
\begin{equation}\label{define_time}
t_\mathrm{S}=t-t_0.
\end{equation}
As we are interested in dynamic events near the saddle point and
the initial site we introduce difference vectors relative to the
position of the saddle point and the initial site, respectively,
i.e.
\begin{equation}\label{define_r_S}
\vec{r}_\mathrm{S}=\vec{r}-\vec{r}_\mathrm{sl}
\end{equation}
and
\begin{equation}\label{define_r_I}
\vec{r}_\mathrm{I}=\vec{r}-\vec{r}_\mathrm{init}.
\end{equation}
Obviously, one could also define a position vector
$\vec{r}_\mathrm{E}$ relative  to the position of the ion site
after the hop, but due to time reversal symmetry, this would not
lead to any new information.

It seems to be most instructive to describe the position vectors $\vec{r}_\mathrm{S}$ and $\vec{r}_\mathrm{I}$
in polar coordinates:
\begin{equation}
\vec{r}_\mathrm{S}=\left(\begin{array}{c}r_\mathrm{S}\\ \theta_\mathrm{S}\\ \phi_\mathrm{S} \end{array}\right) ,\qquad
\vec{r}_\mathrm{I}=\left(\begin{array}{c}r_\mathrm{I}\\ \theta_\mathrm{I}\\ \phi_\mathrm{I} \end{array}\right).
\end{equation}
Here, the angle between the hopping vector
$\vec{r}_\mathrm{end}-\vec{r}_\mathrm{init}$ and the particular
position vector is denoted by $\theta_\mathrm{S}$ and
$\theta_\mathrm{I}$, respectively. $\phi_\mathrm{S}$ and
$\phi_\mathrm{I}$ describe the angle in the plane perpendicular to
the hopping vector. For reasons of isotropy no dependence on
$\phi_\mathrm{S}$ or $\phi_\mathrm{I}$ is expected. The
definitions are visualized in fig.\ \ref{frame}.

\subsection{Definition of the investigated density functions}
\label{define_functions} Let $\left< \rho_i\left(r_\mathrm{S},
\theta_\mathrm{S}, t_\mathrm{S}\right)\right>$ denote the number
density of particles at time $t_\mathrm{S}$ belonging to species
$i$ inside a small volume element characterized by $r_\mathrm{S}$
and $\theta_\mathrm{S}$ and averaged over all identified lithium
hops. This quantity has been investigated for the hopping lithium
ion (HLi) and the non-bridging oxygens (NBO). Actually, we also
analyzed the bridging oxygens and the adjacent lithium ions, but
these results will not be shown explicitly in this paper. By
relating the values of $\left< \rho_i\left(r_\mathrm{S},
\theta_\mathrm{S}, t_\mathrm{S}\right)\right>$ to the total number
density $\bar{\rho}_\mathrm{Li}$ of lithium or
$\bar{\rho}_\mathrm{O}$ of oxygen particles, respectively, one
ends up with:
\begin{equation}\label{define_rel_dens}
\begin{array}{rcl}
\mathrm{P}_\mathrm{HLi}(r_\mathrm{S},\theta_\mathrm{S},t_\mathrm{S})&=&\left<\rho_\mathrm{HLi}(r_\mathrm{S},\theta_\mathrm{S},t_\mathrm{S})\right> /\bar{\rho}_\mathrm{Li}\, ,\vspace{3mm}\\
\mathrm{P}_\mathrm{NBO}(r_\mathrm{S},\theta_\mathrm{S},t_\mathrm{S})&=&\left<\rho_\mathrm{NBO}(r_\mathrm{S},\theta_\mathrm{S},t_\mathrm{S})\right>
/\bar{\rho}_\mathrm{O}\, .
\end{array}
\end{equation}

\section{Results}
\label{Results}
\subsection{Lithium dynamics}
\label{LiDyn} The quantity
$\mathrm{P}_\mathrm{HLi}(r_\mathrm{S},\theta_\mathrm{S},t_\mathrm{S})$
is plotted in fig.\ \ref{Rho_HLi_pol} for five different times
$-2\;\mathrm{ps}\leq t_\mathrm{S}\leq 0$. By definition the
contributing ions perform a hop from the left to the right. The
time evolution is as follows. 1.) At
$t_\mathrm{S}=-2\;\mathrm{ps}$ $\mathrm{P}_\mathrm{HLi}$ shows a
pronounced maximum at the initial site around $r_\mathrm{S}\approx
1.2\;\mathrm{\AA}$ and $\theta_\mathrm{S}=\pi$. From there an area
of high density of hopping lithium spans the saddle along the axis
$\theta_\mathrm{S}=0$ and reaches into the new site up to
$r_\mathrm{S}\approx 1\;\mathrm{\AA}$. This reflects the presence
of lithium ions which are about to hop from the site at
$\vec{r}_\mathrm{end}$ to $\vec{r}_\mathrm{init}$ and, of course,
will jump back at $t_\mathrm{S} = 0$. Thus this intensity reflects
back-and-forth dynamics.  2.) At $t_\mathrm{S}=-500\;\mathrm{fs}$
the area where the density of hopping lithium is finite cumulates
around the initial site. The maximum of $\mathrm{P}_\mathrm{HLi}$
has moved into the direction of the saddle and is now located at
$r_\mathrm{S}\approx 1\;\mathrm{\AA}$ and $\theta_\mathrm{S}=\pi$.
3.) At $t_\mathrm{S}=-120\;\mathrm{fs}$ $\mathrm{P}_\mathrm{HLi}$
shows a broad maximum along the direction of the hopping vector
between the initial site and the saddle point with center at
$r_\mathrm{S}\approx 0.7\;\mathrm{\AA}$ and
$\theta_\mathrm{S}=\pi$. The density is strongly concentrated
around this maximum. 4.) At $t_\mathrm{S}=-40\;\mathrm{fs}$ the
maximum concentration is close to  $r_\mathrm{S}\approx
0.4\;\mathrm{\AA}$ and $\theta_\mathrm{S}=\pi$. 5.) Finally, at
$t_\mathrm{S}=0$ the density is radially distributed around the
saddle point. It is maximal within a radius $r_\mathrm{S}\approx
0.3\;\mathrm{\AA}$ and decays rapidly for larger values of
$r_\mathrm{S}$. Due to the fact that the definition of
$\vec{r}_\mathrm{sl}$ involves an average over several time steps,
finite values of $r_S$ are possible for $t_0 = 0$.

Of course, the cumulation of the density of hopping lithium while
$t_\mathrm{S}$ approaches zero is due to the fact, that by
definition an ion is crossing the saddle close to
$t_\mathrm{S}=0$. The broadening at earlier times shows that
different ions behave quite differently in the time interval just
before the hop.

Emphasizing the effect of highly occupied lithium sites, it has
turned out that at least for a short time during the shown time
interval $-2\;\mathrm{ps}\leq t_\mathrm{S}\leq 0$ the site close
to $\vec{r}_\mathrm{end}$ is occupied by another ion for about
$83\;\%$ of all identified lithium hops. This does not only imply
a high probability to find an ion at a site almost immediately
before another ion will perform a hop to this site. It also means
a high probability for a site to be occupied again by an ion just
after another ion has hopped out of it. We found the time scale of
this re-occupation to increase slightly with decreasing
temperature, but even at the significantly lower temperature
$T=640\;\mathrm{K}$ more then $65\;\%$ of the initial sites have
been re-occupied by another ion at $t_\mathrm{S}=2\;\mathrm{ps}$
(not shown).

\subsection{Oxygen dynamics around the saddle}
\label{ODyn} Before we focus on the oxygen dynamics during an ion
hop, we show the partial pair correlation functions
$g_\mathrm{Li\,NBO}(r)$ for lithium and non-bridging oxygen (fig.\
\ref{g_r_Li_NBO_BO}). $g_\mathrm{Li\,NBO}(r)$ is different from
zero for distances $r\gtrsim 1.6\;\mathrm{\AA}$. A pronounced
maximum is found at $r\approx 2.0\;\mathrm{\AA}$ representing the
first shell of non-bridging oxygen around lithium. The maxima for
the second and higher order shells of non-bridging oxygen are less
pronounced.

We can now compare this distance with characteristic distances
$r_\mathrm{S}$ of
$\mathrm{P}_\mathrm{NBO}(r_\mathrm{S},\theta_\mathrm{S},t_\mathrm{S})$.
The time evolution of $\mathrm{P}_\mathrm{NBO}$ is shown in fig.\
\ref{Rho_NBO_pol}. 1.) At $t_\mathrm{S}=-2\;\mathrm{ps}$ the
density of non-bridging oxygen indicates two minima around the
centers of the two lithium sites. The minimum around the initial
site seems to be a little more pronounced. This matches the
slightly but clearly higher lithium density at this site.
$\mathrm{P}_\mathrm{NBO}$ shows a broad maximum around
$r_\mathrm{S}\approx 1.8\;\mathrm{\AA}$ and
$\theta_\mathrm{S}\approx\pi /2$. The shortest distance to the
saddle where $\mathrm{P}_\mathrm{NBO}$ reaches zero is at
$r_\mathrm{S}\approx 0.6\;\mathrm{\AA}$. 2.) While time approaches
zero the minima in non-bridging oxygen density get more and more
pronounced. At the same time the density spike for
$\theta_\mathrm{S}\approx\pi /2$ separating the two minima
vanishes. The maximum of $\mathrm{P}_\mathrm{NBO}$ broadens in
$\theta_\mathrm{S}$ direction and narrows in $r_\mathrm{S}$
direction. Its peak increases and moves to slightly higher values
of $r_\mathrm{S}$. At $t_\mathrm{S}=0$ nearly everywhere inside an
approximately spherical volume with radius $r_\mathrm{S}\approx
1.5\;\mathrm{\AA}$ around the saddle the density of non-bridging
oxygen has vanished. The maximum is shifted to
$r_\mathrm{S}\approx 1.9\;\mathrm{\AA}$. A similar behavior is
observed for the bridging oxygens.

The time evolution of
$\mathrm{P}_\mathrm{NBO}(r_\mathrm{S},\theta_\mathrm{S},t_\mathrm{S})$
gives evidence that during an ion hop the oxygen particles around
the saddle point significantly retreat. This mechanism can be
compared to a sliding door which has to be opened to let the ion
pass. To investigate this effect in more detail, we show another
plot of $\mathrm{P}_\mathrm{NBO}$,
in which we concentrate on a thin slice of $0.5\;\mathrm{\AA}$
thickness perpendicular to the hopping vector around
$r_\mathrm{S}=0$; see again fig.\ \ref{frame}. The results are
shown in fig.\ \ref{Rho_NBO_peak}. We have also included the
corresponding data for a system with an immobilized network at the
same temperature $T=980\;\mathrm{K}$.

As the most pronounced effect oxygen density for $r_\mathrm{S} <
1.5\;\mathrm{\AA}$ is dramatically reduced when $t_\mathrm{S}$
approaches zero. In case of the immobilized network the density of
NBOs is also negligible for $r_\mathrm{S} < 1.5\;\mathrm{\AA}$.
Fig.\ \ref{Rho_NBO_peak} again shows the sliding door mechanism as
already mentioned in the discussion of fig.\ \ref{Rho_NBO_pol}. It
seems to be related to a critical minimum distance between oxygen
and the saddle at the time it is crossed by the hopping ion. This
is $r_\mathrm{S}\approx 1.5\;\mathrm{\AA}$ for non-bridging
oxygen. Very close to this value the respective pair correlation
functions decays to zero (see fig.\ \ref{g_r_Li_NBO_BO}). The
remaining small differences do not necessarily mean that during a
hop the distance between lithium and non-bridging oxygen falls
below $1.6\;\mathrm{\AA}$, because the ion does not have to
permeate the plane around the saddle strictly at
$\vec{r}_\mathrm{S}=0$.

To investigate the sliding door mechanism somewhat closer we
analyze the distribution of non-briding oxygen atoms at
$t_\mathrm{S}=0$ for different subensembles. Subensemble $A_1$ is
defined as the set of oxygens with distances $r_\mathrm{S}\leq
1.5\;\mathrm{\AA}$ from a saddle at time
$t_\mathrm{S}=-1\;\mathrm{ps}$. Subensemble $A_2$ is defined via
the same criterion, but at time $t_\mathrm{S}=-2\;\mathrm{ps}$.
 The
oxygen-distribution of subensemble $A_2$ at time
$t_\mathrm{S}=-1\;\mathrm{ps}$ is distributed around
$r_\mathrm{s}\approx 1.5\;\mathrm{\AA}$ and is therefore very
different to subensemble $A_1$ at that time ($r_\mathrm{S}\leq
1.5\;\mathrm{\AA}$); see fig.\ \ref{comp_sub}. This difference
reflects the typical fluctuations of the oxygen atoms during 1 ps
{\it without} an ion hop. Now the interesting question emerges
whether the oxygen distribution at $t_\mathrm{S}=0\;\mathrm{ps}$
is different for both subensembles. As also shown in fig.\
\ref{comp_sub} both distributions are identical. Comparison with
$\mathrm{P}_\mathrm{NBO}(r_\mathrm{S}, t_\mathrm{S})$ for all
non-bridging oxygens at time $t_\mathrm{S}=0$ shows that this
distribution is very similar to the overall distribution at
$t_\mathrm{S}=0$ after appropriate scaling to the number of
contributing particles. Thus on average the position of oxygens
close to the saddle at time $t_\mathrm{S}=0$, i.e. the time of the
lithium hop, does not depend on its position in the past.
Furthermore it turns out that the oxygen distribution
$\mathrm{P}_\mathrm{NBO}(r_\mathrm{S}, t_\mathrm{S}=0)$ only
weakly depends on temperature, as shown from the comparison of
$T=980\;\mathrm{K}$ K with $T=640\;\mathrm{K}$. The implications
of these observations will be discussed below.

\subsection{Oxygen dynamics around the initial site}
\label{ODyn2} In fig.\ \ref{Rho_NBO_site} we show the time
evolution of the relative density
$\mathrm{P}_\mathrm{NBO}(r_\mathrm{I},t_\mathrm{S})$ of
non-bridging oxygen inside a hemisphere around
$\vec{r}_\mathrm{init}$ covering angles $\pi /2
\leq\theta_\mathrm{I}\leq\pi $. At time
$t_\mathrm{S}=-1\;\mathrm{ps}$ the first maximum for
$\mathrm{P}_\mathrm{NBO}(r_\mathrm{I},t_\mathrm{S})$ is found at
$r_\mathrm{I}\approx 2.1\;\mathrm{\AA}$. The density decays to
zero at $r_\mathrm{I}\approx 1.4\;\mathrm{\AA}$. When time
$t_\mathrm{s}$ approaches zero, i.e. the ion leaves the initial
site, the corresponding peak gets broader and slightly shifts to
larger distances.  This trend stops around $t_\mathrm{S}\approx
120$ fs. For larger times the density tends to approach the
initial density before the jump. The reason is that very likely
the initial site will be populated by another ion. In summary, two
major effects result after vacating an ion site. First, the
network structures somewhat loosens as reflected by the broadening
of the distribution and, second, the surrounding oxygen atoms
slightly increase their distance from this site.

\section{Conclusions}
\label{concl} In this paper we have introduced a useful tool to
identify  the time and the position of ion hops. This enables us
to study the temporal and spatial evolution of the relative
density distribution of hopping ions and adjacent oxygen atoms.
This leads to an instructive picture of the microscopic ion
hopping dynamics. It turns out that there is a high probability
for a site to be re-occupied by an ion almost immediately after it
has been left. Even for the lowest investigated temperature
$T=640\;\mathrm{K}$ this re-occupation takes place on a time scale
of picoseconds. This behavior is similar to results obtained for a
sodium silicate system by Cormack \textit{et al.}
\cite{cormack02}. It also matches the perception of Habasaki
\textit{et al.} \cite{habasaki95-2, habasaki96, habasaki97,
habasaki99}.

We have shown that oxygen atoms near the saddle point of an ion
hop retreat when the hop occurs. The oxygen motion reminds of a
sliding door that opens to let the ion pass. Here a minimum
distance to the saddle is kept by the oxygen which is very close
to the distance where the corresponding pair correlation function
with lithium vanishes.

This cooperative dynamics between lithium and those oxygen atoms
close to the saddle can be easily represented in an effective
2D-energy landscape of the type as sketched in fig.\
\ref{energylandscape}. Of course, in reality the energy landscape
is 3N-dimensional (N: particle number). The common reaction
coordinate contains the lithium hop as well as the back-and-forth
dynamics of the oxygen. Of course, for different saddles and
different oxygen atoms the specific energy landscape may somewhat
look different. At $t_\mathrm{S}=0$ one may define an effective
oxygen potential $V_i(R_\mathrm{O})$. The index $i$ counts the
different effective potentials, experienced by the different
oxygen atoms. Now two length scales can be determined.
$\xi_\mathrm{hom}$ denotes the homogeneous length scale describing
the range of possible fluctuations in the potential
$V_i(R_\mathrm{O})$. Close to the minima one expects in harmonic
approximation $\xi_\mathrm{hom} \propto T^{1/2}$. In contrast,
$\xi_\mathrm{het}$ accounts for the heterogeneity when comparing
hopping processes at different locations or different oxygen
atoms. More specifically, it denotes the range of positions for
which the different potentials $V_i(R_\mathrm{O})$ display their
respective minima. The two extreme limits $\xi_\mathrm{het} \gg
\xi_\mathrm{hom}$ and vice versa are sketched in fig.\
\ref{V_i_r}. To elucidate the ratio of both length scales in the
present case we would like to repeat two important observations,
presented above. First, for oxygen atoms close to the saddle the
position at $t_\mathrm{S} = 0$ does not depend on its position 1
ps before the hop as derived from the comparison of the behavior
of two different subensembles. Second, the oxygen distribution at
$t_\mathrm{S}=0$ only shows an extremely weak temperature
dependence. Both observations are compatible with the limit
$\xi_\mathrm{het} \gg \xi_\mathrm{hom}$. In this case the position
of any oxygen at $t_\mathrm{S}=0$ is very well characterized by
its effective potential $V_i(R_\mathrm{O})$ and does (in the
extreme limit $\xi_\mathrm{hom} \rightarrow 0)$ not depend on its
previous positions. Furthermore, since the effective energy
landscape is to a large extent temperature independent one would
not expect a major temperature dependence of the oxygen
distribution, in agreement with observation. In any event, one may
estimate $\xi_\mathrm{hom} \ll 0.5\; \mathrm{\AA}$, i.e. the width
of $\mathrm{P}_\mathrm{NBO}(r_\mathrm{S}, t_\mathrm{S}=0)$.

These results also reveal possible mechanisms for the dramatic
slowing down of the ion dynamics upon immobilization of the
network \cite{heuer05}. From the  fig.\ \ref{comp_sub} one can see
that  many oxygen atoms have to move as much as
$0.5\;\mathrm{\AA}$ to arrive at their optimum position at
$t_\mathrm{S}=0$. In case of an immobilized network this motion is
not possible. This means that many oxygen atoms may be much
further away from their optimum position during the transition as
compared to $\xi_\mathrm{hom}$. This automatically implies that
the effective barrier for the combined lithium-oxygen transition
will be much higher. This effect is sketched in fig.\
\ref{energylandscape} where the suppression of the oxygen dynamics
will lead to a significantly higher barrier of the transition in
probably most of the hopping events. Of course, if the oxygen
distance to a saddle is less than  $1.5\;\mathrm{\AA}$ the
resulting saddle is so high that effectively no lithium hopping is
possible along this path. One can estimate from the oxygen
distribution in fig.\ \ref{Rho_NBO_peak} at $t_\mathrm{S}=-2$ ps
that this is the case for roughly 10\% of all saddles. In any
event, based on these considerations the dramatic increase of the
activation energy upon immobilization can be understood.

 As a gedankenexperiment one may
wonder for which (unphysical) oxygen mass one has the transition
between the mobile and the immobile limit. Based on the above
interpretation of the correlated hopping motion it is required
that the oxygen can follow the lithium ion during the time of a
hop, which takes of the order of 500 fs. Thus the velocity of the
oxygen atom should be decorrelated on time scales faster than 500
fs. For this purpose we analyze the quantity $w(t)$, which is the
time derivative of the mean square displacement and twice the time
integral of the velocity autocorrelation function \cite{funke02,
heuer02}. It is plotted in fig.\ \ref{w_t}. The short-time peak of
$w(t)$ is a measure for the typical velocity decorrelation time.
It is of the order of 30 fs for oxygen and thus much faster than
the typical jump times of the lithium ions. Thus one can estimate
that by increasing the oxygen mass by $(500/30)^2$ the slowing
down of the lithium dynamics should start.

Actually, a similar estimation has been performed by Sunyer
\textit{et al.} for MD simulations of a sodium silicate system
\cite{sunyer03}. They explicitly observed the slowing down in
dependence of the (artificial) mass of the network atoms. To
rationalize the slowing down they have compared the fourier
transform of the velocity correlation function of the sodium atoms
with that of the oxygen atoms. This quantity expresses the
vibronic density of states. They found that an overlap is
important for the sodium dynamics. As soon as the oxygen atoms
become too slow and the overlap decreases the sodium diffusion
constant starts to decrease, too. This happens if the oxygen mass
is scaled by a factor of the order of 100-1000. This estimation is
close to the present case.

A further conclusion about the interrelation of lithium and
network dynamics, as elucidated in this work, can be drawn from
fig.\ \ref{w_t}. First, it is important to realize that around $t
\approx $ 1 ps there is a transition from localized lithium
dynamics in a single site to hopping dynamics between adjacent
sites. This has been explicitly shown in \cite{heuer02}. This time
scale is also consistent with the present results because the
hopping transition also roughly takes 1 ps. One can directly
observe that up to 1 ps the observable $w(t)$ of the network atoms
only weakly depends on temperature. This is in agreement with the
expectation for the vibrational short-time limit ( $w(t) \propto
T^{1/2})$. Around 1 ps the temperature dependence dramatically
increases and is comparable with the temperature dependence of
$w(t)$ for the lithium ions. Since in this time regime the lithium
ions start to hop this correlation may reflect the sliding door
mechanism as well as the network relaxation at ion sites,
discussed in this work. For these mechanism one indeed expects a
strong relation between network and lithium dynamics.

We just note in passing that the comparison of $w(t)$ for
different species also has implications beyond the immediate topic
of this work. In conductivity experiments one observes the sum of
the transport of all charged constituents of the probe. Their
contribution to $\sigma(\omega)$ is proportional to $w(t \approx
\pi^2/\omega)$ \cite{funke02, heuer02}. Thus one can conclude that
for $t > 1 $ ps, i.e. in the hopping regime, the contribution is
mainly due to lithium dynamics.

In summary, we have elucidated in great detail the interrelation
of lithium and network dynamics. It turns out that the reaction
path for transition involves a significant oxygen dynamics. It can
be characterized with a sliding door mechanism. There are strong
indications that this dynamics is crucial to reduce the barrier
height as compared to the case where the oxygen dynamics is
suppressed due to network immobilization.

\section*{Acknowledgement}
Financial support from the DFG in the framework of SFB 458 is
gratefully acknowledged. We appreciate helpful discussions with R.
Banhatti, C. Cramer-Kellers, K. Funke, M.D. Ingram, H. Lammert, M.
Vogel and D. Wilmer about this topic.

\clearpage

\bibliographystyle{pccp}
\bibliography{slidedoor2_pccp}

\begin{thebibliography}{10}

\bibitem{howell74}
F.~Howell, R.~Bose, P.~Macedo, and C.~Moynihan, {\em J.\ Phys.\ Chem.}, 1974,
  {\bf 78}, 639.

\bibitem{angell86}
C.~Angell, {\em Solid State Ionics}, 1986, {\bf 18-19}, 72.

\bibitem{heuer02}
A.~Heuer, M.~Kunow, M.~Vogel, and R.~Banhatti, {\em Phys.\ Chem.\ Chem.\
  Phys.}, 2002, {\bf 4}, 3185.

\bibitem{angell81}
C.~Angell, L.~Boehm, P.~Cheeseman, and S.~Tamaddon, {\em Solid State Ionics},
  1981, {\bf 5}, 659.

\bibitem{sunyer03}
E.~Sunyer, P.~Jund, and R.~Jullien, {\em J.\ Phys.: Condens.\ Matter}, 2003,
  {\bf 15}, L431.

\bibitem{habasaki04}
J.~Habasaki, K.~Ngai, and Y.~Hiwatari, {\em J.\ Chem.\ Phys.}, 2004, {\bf 120},
  8195.

\bibitem{heuer05}
A.~Heuer, H.~Lammert, and M.~Kunow, {\em Z.\ f.\ Phys.\ Chemie}, 2004, {\bf
  218}, 1429.

\bibitem{anderson54}
O.~Anderson and D.~Stuart, {\em J.\ Am.\ Ceram.\ Soc.}, 1954, {\bf 37}, 537.

\bibitem{martin86}
S.~Martin and C.~Angell, {\em Solid State Ionics}, 1986, {\bf 23}, 185.

\bibitem{cmmc}
P.~Bandaranayake, C.~Imrie, and M.~Ingram, {\em Phys. Chem. Chem. Phys.}, 2002,
  {\bf 4}, 3209.

\bibitem{ingramnew}
M.~Ingram, R.~Banhatti, and I.~Konidakis, {\em Z. Phys. Chem}, 2004, {\bf 218},
  1401.

\bibitem{habasaki_new}
J.~Habasaki, K.~Ngai, and Y.~Hiwatari, {\em J.\ Chem.\ Phys.}, 2004, {\bf 121},
  925.

\bibitem{lammert03}
H.~Lammert, M.~Kunow, and A.~Heuer, {\em Phys.\ Rev.\ Lett.}, 2003, {\bf 90},
  215901.

\bibitem{vogel:2004}
M.~Vogel, {\em Phys. Rev. B}, 2004, {\bf 70}, 139902.

\bibitem{horbach:2002}
J.~Horbach, W.~Kob, and K.~Binder, {\em Phys. Rev. Lett.}, 2002, {\bf 88},
  125502.

\bibitem{cormack02}
A.~Cormack, J.~Du, and T.~Zeitler, {\em Phys.\ Chem.\ Chem.\ Phys.}, 2002, {\bf
  4}, 3193.

\bibitem{habasaki95-2}
J.~Habasaki, I.~Okada, and Y.~Hiwatari, {\em Phys.\ Rev.\ E}, 1995, {\bf 52},
  2681.

\bibitem{habasaki96}
J.~Habasaki, I.~Okada, and Y.~Hiwatari, {\em J.\ Non-Cryst.\ Solids}, 1996,
  {\bf 208}, 181.

\bibitem{habasaki97}
J.~Habasaki, I.~Okada, and Y.~Hiwatari, {\em Phys.\ Rev.\ B}, 1997, {\bf 55},
  6309.

\bibitem{habasaki99}
J.~Habasaki and Y.~Hiwatari, {\em Phys.\ Rev.\ E}, 1999, {\bf 59}, 6962.

\bibitem{banhatti01}
R.~Banhatti and A.~Heuer, {\em Phys.\ Chem.\ Chem.\ Phys.}, 2001, {\bf 3},
  5104.

\bibitem{habasaki90}
J.~Habasaki, {\em Mol.\ Phys.}, 1990, {\bf 70}, 513.

\bibitem{habasaki92}
J.~Habasaki and I.~Okada, {\em Mol.\ Simul.}, 1992, {\bf 9}, 319.

\bibitem{habasaki93}
J.~Habasaki, I.~Okada, and Y.~Hiwatari, {\em Mol.\ Simul.}, 1993, {\bf 10}, 19.

\bibitem{refson}
K.~Refson, {\em Comput.\ Phys.\ Commun.}, 2000, {\bf 126}, 310.

\bibitem{doweidar}
H.~Doweidar, {\em J.\ Non-Cryst.\ Solids}, 1996, {\bf 194}, 155.

\bibitem{funke02}
K.~Funke, R.~Banhatti, S.~Brückner, C.~Cramer, C.~Krieger, A.~Mandanici,
  C.~Martiny, and I.~Ross, {\em Phys.\ Chem.\ Chem.\ Phys.}, 2002, {\bf 4},
  3155.

\end{thebibliography}

\begin{figure}[ht]
  \begin{center}
   \includegraphics[width=0.63\linewidth]{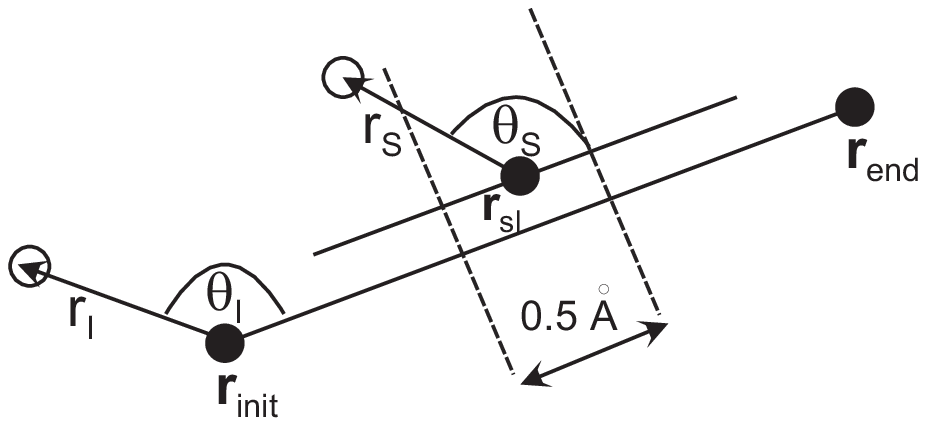}
   \caption{\label{frame}}
  \end{center}
\end{figure}

\begin{figure}[ht]
  \begin{center}
   \includegraphics[width=\linewidth]{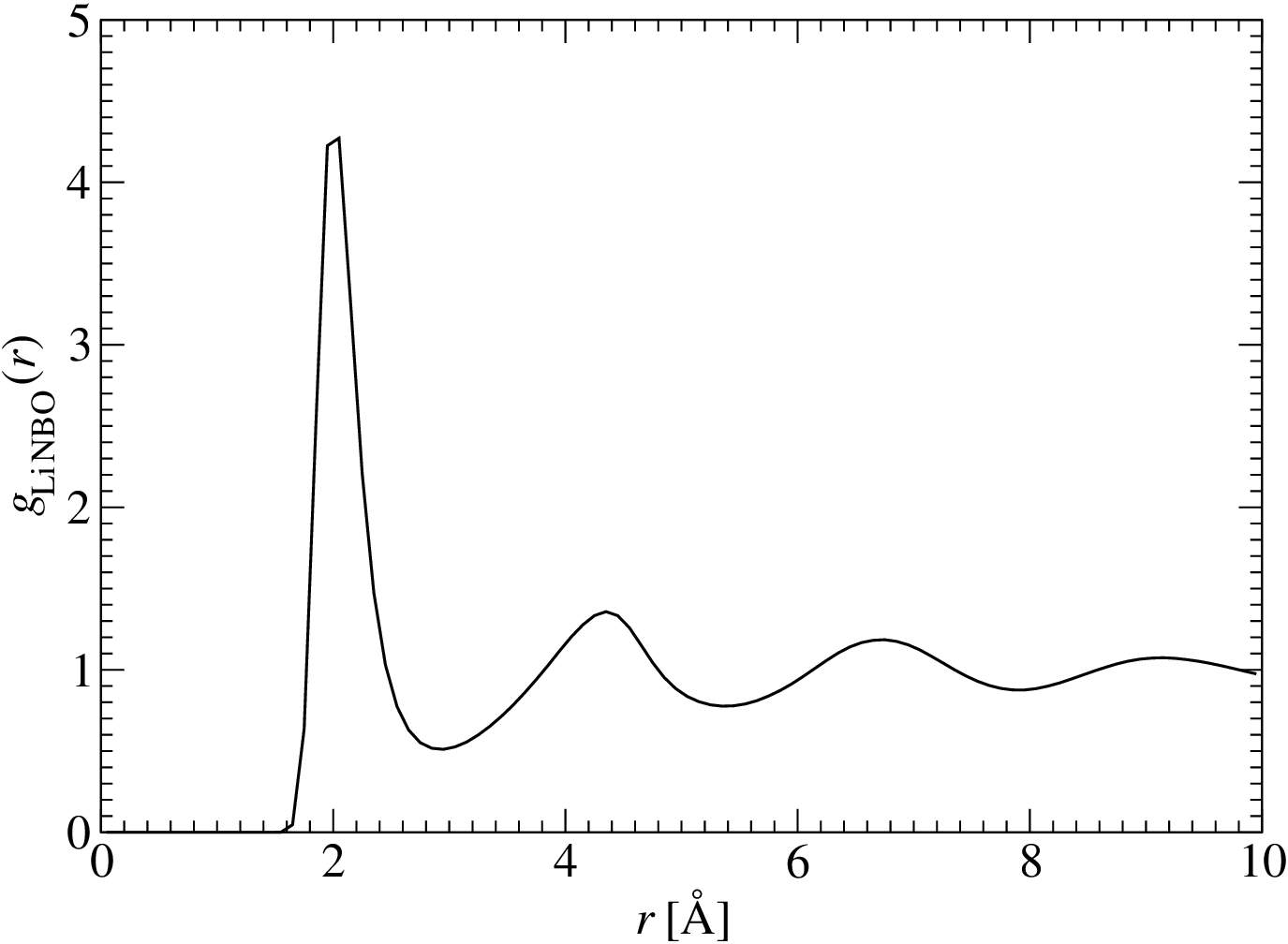}
   \caption{\label{g_r_Li_NBO_BO}}
  \end{center}
\end{figure}

\begin{figure}[ht]
  \begin{center}
   \includegraphics[width=0.43\linewidth]{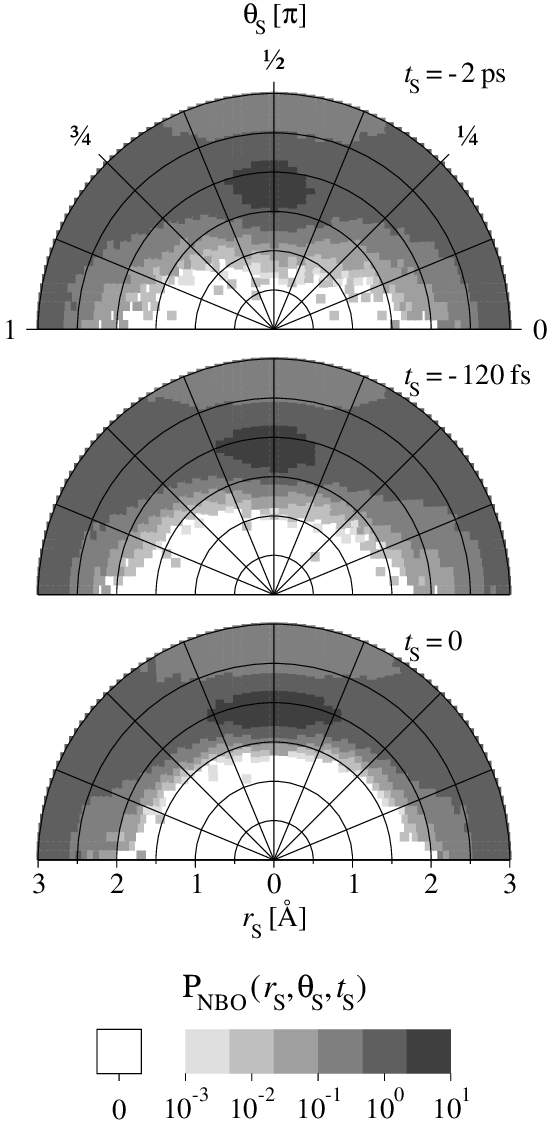}
   \caption{\label{Rho_NBO_pol}}
  \end{center}
\end{figure}

\begin{figure}[ht]
  \begin{center}
   \includegraphics[width=\linewidth]{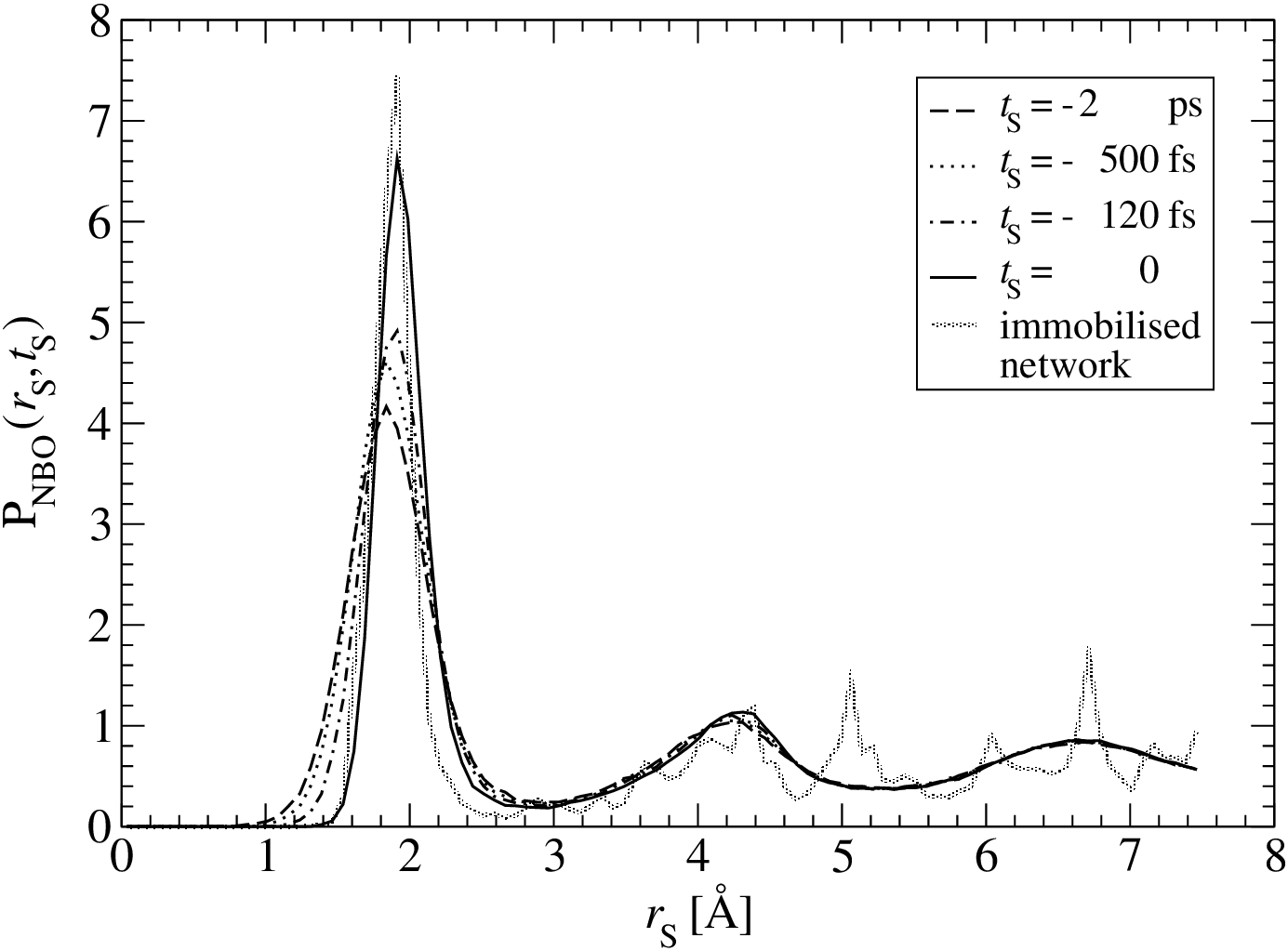}
   \caption{\label{Rho_NBO_peak}}
  \end{center}
\end{figure}

\begin{figure}[ht]
  \begin{center}
   \includegraphics[width=\linewidth]{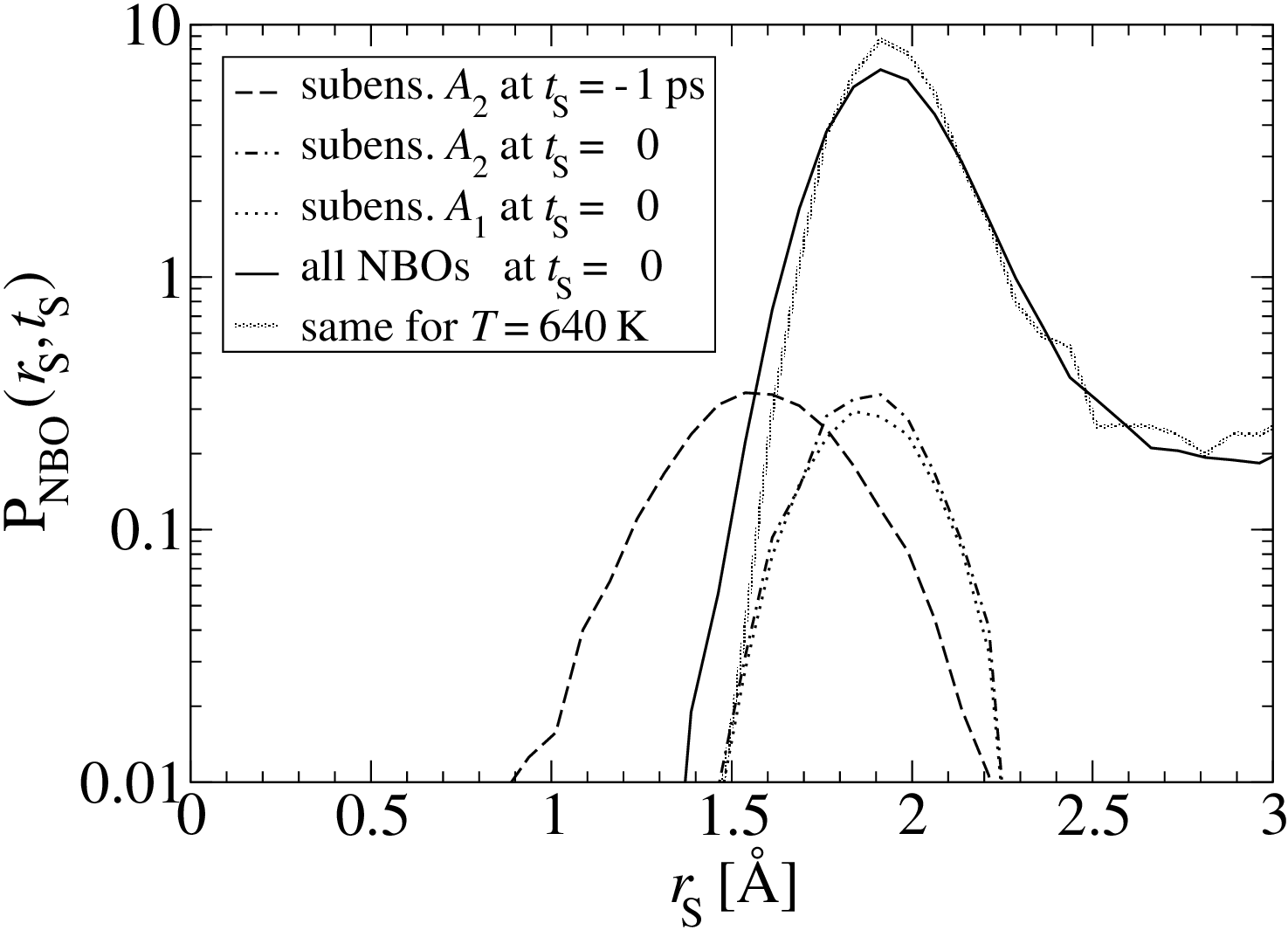}
   \caption{\label{comp_sub}}
  \end{center}
\end{figure}

\begin{figure}[ht]
  \begin{center}
   \includegraphics[width=\linewidth]{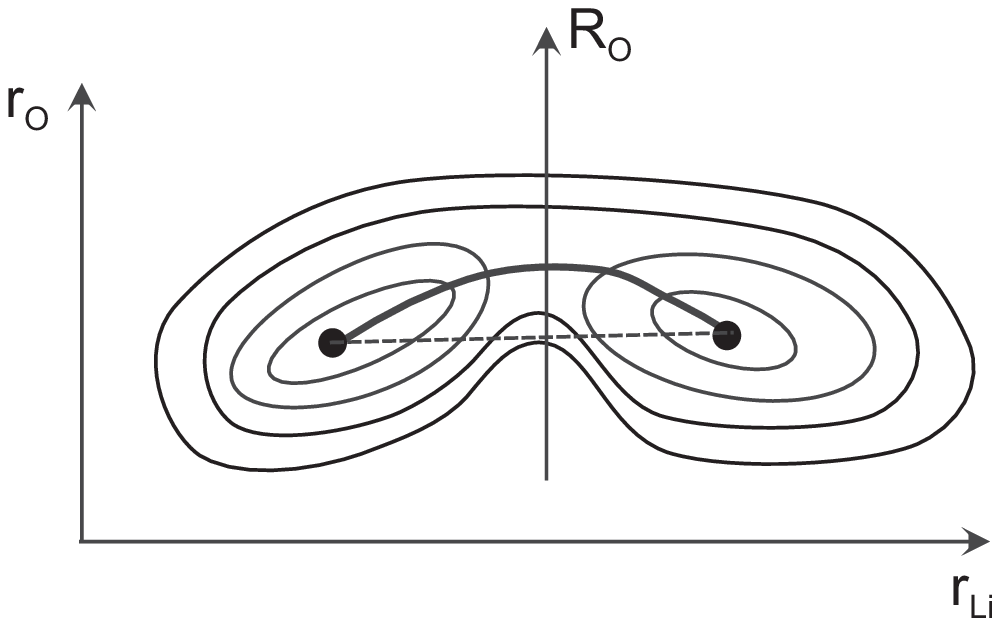}
   \caption{\label{energylandscape}}
  \end{center}
\end{figure}

\begin{figure}[ht]
  \begin{center}
   \includegraphics[width=\linewidth]{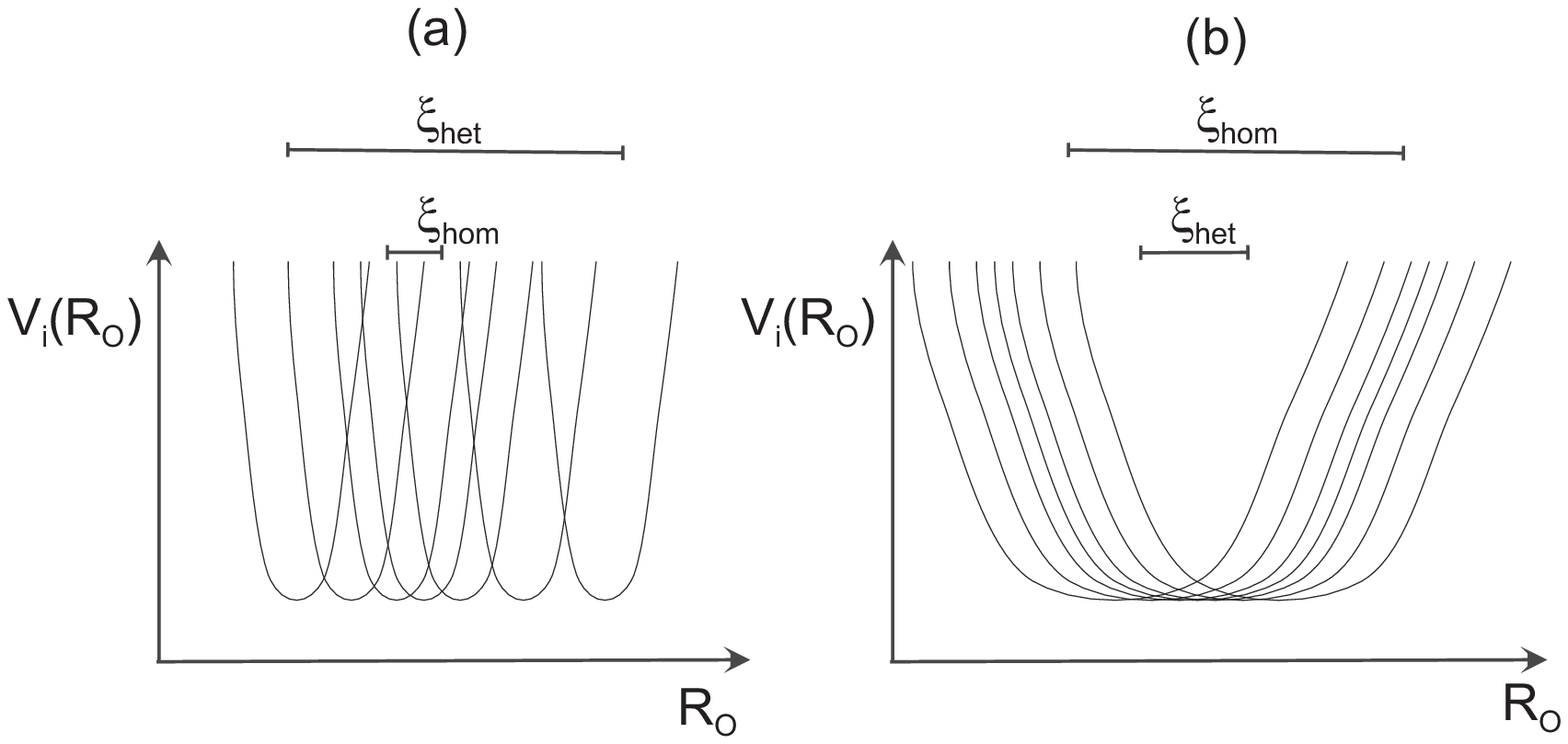}
   \caption{\label{V_i_r}}
  \end{center}
\end{figure}

\clearpage
\begin{list}{}{\leftmargin 2cm \labelwidth 1.5cm \labelsep 0.5cm}

\item[\bf Fig. \ref{frame}] Definition of the coordinate frame
used in this work. The thin slice around $r_{sl}$ will be used
further below.

\item[\bf Fig. \ref{Rho_HLi_pol}] Temporal and spatial evolution
of the relative density $\mathrm{P}_\mathrm{HLi}\left(
r_\mathrm{S}, \theta_\mathrm{S}, t_\mathrm{S}\right)$ of hopping
lithium.

\item[\bf Fig. \ref{g_r_Li_NBO_BO}] Pair correlation function
$g_{\mathrm{Li}}(r)$ for lithium and non-bridging oxygen
($i=\mathrm{NBO}$).

\item[\bf Fig. \ref{Rho_NBO_pol}] Temporal and spatial evolution
evolution of the relative density $\mathrm{P}_\mathrm{NBO}\left(
r_\mathrm{S}, \theta_\mathrm{S}, t_\mathrm{S}\right)$ of
non-bridging oxygen adjacent to hopping lithium.

\item[\bf Fig. \ref{Rho_NBO_peak}] Temporal and spatial evolution
of the relative density $\mathrm{P}_\mathrm{NBO}\left(
r_\mathrm{S}, t_\mathrm{S}\right)$ of non-bridging oxygen in a
$0.5\;\mathrm{\AA}$ thin slice around the plane
$\theta_\mathrm{S}=\pi /2$.

\item[\bf Fig. \ref{comp_sub}] Relative density
$\mathrm{P}_\mathrm{NBO}\left( r_\mathrm{S}, t_\mathrm{S}\right)$
of non-bridging oxygen in a $0.5\;\mathrm{\AA}$ thin slice around
the plane $\theta_\mathrm{S}=\pi /2$ for different subensembles
$A_1$ and $A_2$ compared to $\mathrm{P}_\mathrm{NBO}\left(
r_\mathrm{S}, t_\mathrm{S}\right)$ for all non-bridging oxygens in
this slice. Particles belonging to subensemble $A_1$ are located
at $r_\mathrm{S}\leq 1.5\;\mathrm{\AA}$ at
$t_\mathrm{S}=-1\;\mathrm{ps}$. Particles belonging to subensemble
$A_2$ are located at $r_\mathrm{S}\leq 1.5\;\mathrm{\AA}$ at
$t_\mathrm{S}=-2\;\mathrm{ps}$.

\item[\bf Fig. \ref{Rho_NBO_site}] Temporal and spatial evolution
of the relative density $\mathrm{P}_\mathrm{NBO}\left(
r_\mathrm{I}, t_\mathrm{S}\right)$ of non-bridging oxygen in a
hemisphere around $\vec{r}_\mathrm{init}$ covering angles $\pi /2
\leq \theta_\mathrm{I}\leq\pi$.

\item[\bf Fig. \ref{energylandscape}] Schematic 2D energy
landscape, reflecting the cooperative motion of the lithium and an
oxygen atom. The thick line describes the reaction coordinate. The
straight line for constant $r_O$ describes the transition path
without oxygen dynamics. In contrast, along the straight line for
constant $r_\mathrm{Li}$ one can read off the effective potential
$V_i(R_\mathrm{O})$for the oxygen atom at $t_\mathrm{S} = 0$.

\item[\bf Fig. \ref{V_i_r}] Sketch of the two extreme scenario of
the effective potentials $V_i(R_\mathrm{O})$, experienced by the
oxygen atoms at time $t_\mathrm{S}=0$. (a) Limit of strong
curvature; (b) Limit of weak curvature.

\item[\bf Fig. \ref{w_t}] Time derivative $w(t)$ of the mean square displacement for all particle species in the system at three different temperatures.

\end{list}

\clearpage

\begin{figure}[ht]
  \begin{center}
   \includegraphics[width=0.43\linewidth]{Rho_HLi_pol.eps}
   \caption{\label{Rho_HLi_pol}}
  \end{center}
\end{figure}

\begin{figure}[ht]
  \begin{center}
   \includegraphics[width=\linewidth]{Rho_NBO_site.eps}
   \caption{\label{Rho_NBO_site}}
  \end{center}
\end{figure}

\begin{figure}[ht]
  \begin{center}
   \includegraphics[width=\linewidth]{w_t.eps}
   \caption{\label{w_t}}
  \end{center}
\end{figure}

\end{document}